# Quasi-ballistic electron transport in double-wall carbon nanotubes

Hisashi Kajiura [a,b,*], Houjin Huang [b], and Alexey Bezryadin [a]

[a] *Department of Physics, University of Illinois at Urbana-Champaign, 1110 W Green Street, Urbana, Illinois 61801, USA*

[b] *Materials Laboratories, Sony Corporation, 4-16-1 Okata, Atsugi City, Kanagawa 243-0021, Japan*

\* Corresponding author: Hisashi Kajiura

Postal address: Materials Laboratories, Sony Corporation, 4-16-1 Okata, Atsugi City, Kanagawa 243-0021, Japan,

Tel: +81-46-226-3773,

Fax: +81-45-226-2453,

E-mail: hisashi.kajiura@jp.sony.com

ABSTRACT

Room-temperature quasi-ballistic electron transport in double-wall carbon nanotubes (DWNT) is demonstrated. Conductance dependence on the length was measured by submerging DWNTs into liquid mercury. The conductance plots show plateaus, indicating weak dependence of the electrode-tube-electrode electrical resistance on the length of the connecting nanotube. We infer a mean free path between 0.6 and 10 μm for ~80% of the DWNTs, which is in good agreement with calculations based on the electron scattering by acoustic phonons and by disorder.

Double-wall carbon nanotubes (DWNTs) [1] have recently attracted much attention because of their unique structure and theoretically predicted peculiar physical and chemical properties, not found in either single-wall carbon nanotubes or multi-wall carbon nanotubes (MWNTs) [2,3,4,5]. However, due to the difficulties in producing DWNT samples, their fundamental characteristics, such as ballistic transport properties [6], have not been well established [7,8,9,10]. In order to investigate electronic properties of DWNTs, we have developed a new synthesis method [11], which enables us to prepare suitable DWNT samples for transport measurements. Using these DWNT samples, we studied their ballistic transport properties by submerging them into liquid mercury (Hg), following the method pioneered by the de Heer group [12]. The experiments were carried out at room temperature. The results show that the electrons in DWNTs can propagate without scattering over a distance of a few microns, which is comparable to the typical length of our nanotubes. We compare these results to the calculated mean free path limited by scattering of electrons by acoustic phonons [13] and find a good agreement with the measured values. Some tubes showed a shorter mean free path, compared to what is expected if only electron-phonon scattering is involved. In these tubes, there may be some positional disorder of as much as ~2%.

DWNT samples, which appear as threadlike soot, were produced on the surface of a bowl-shaped cathode using arc-discharge in helium (purity >99.9999%) [11]. An image of the as-produced DWNT soot sample obtained with a scanning electron microscopy (SEM) Hitachi-S4700 is shown in Fig. 1A. The micrograph shows DWNT bundles protruding at the edges. Images made using a transmission electron microscope (TEM) Hitachi-HF2000 show DWNTs with the outer diameter in the range 2-7 nm (Fig. 1B). The as-produced soot sample contains catalytic metal particles up to 60% [11].

Transport measurements were made using a piezo-driven nanopositioning system (Fig. 2A). A piezo-positioner allowed gentle and reproducible contact between the nanotube sample and the Hg electrode [12]. The as-produced soot sample was attached to a mobile metallic electrode ("probe") using conducting silver paste. The probe was then attached to a piezo-positioner with a displacement range of 20 μm (17PAZ005, MELLES GRIOT). The liquid Hg electrode was positioned below the mobile electrode (i.e. below the probe), within the displacement range of the piezo-positioner. To make electrical contact between the sample and Hg, the probe was driven cyclically up and down with a peak-to-peak amplitude of 2 – 10 μm and a frequency of 0.5 – 1 Hz. A potential of 50 mV was applied between the probe and Hg electrodes. The current was measured as a function of the piezo positioner displacement with a typical sampling rate of 1000 points/s using an analog-to-digital converter (NI6120, National Instruments). All measurements were carried out at room temperature in air. Following Lovall *et al*. [14], we assume that a single nanotube frequently protrudes from the tip of the bundle (Fig. 2B), and that our measurements provide information about a single nanotube at the end of the bundle. In order to minimize the effect of impurities such as catalytic metal particles, the sample was dipped into mercury more than 1000 times before making measurements [15].

Figure 3A shows a conductance trace versus mobile probe displacement, $G(x)$, normalized by the conductance quantum, $G_0 \equiv 2e^2/h = 7.75 \times 10^{-5}$ S $= (12.9$ k$\Omega)^{-1}$ ($e$ is the elementary charge and $h$ is Planck's constant). Here $x$ represents the piezo positioner extension, with $x = 0$ corresponding to the point at which the tube-Hg contact is established. Thus $x$ gives a measure for the nanotube segment submerged into the Hg electrode (Fig. 2B). Note also that the length of the nanotube segment connecting the probe and the Hg electrodes (i.e. the segment that is not submerged into mercury) equals $L$-$x$, where $L$ is the total length of the nanotube. In the great majority of our measurements, we observed a sequence of steps on the $G(x)$ curves (Fig. 3A). We interpret each step as being caused by a new nanotube (possibly from the same bundle) touching Hg [12]. It is interesting to note that the plateau after each step indicates that the conductance depends very weakly on the length of the nanotube segment connecting the probe and the Hg electrodes. If a *diffusive* metallic wire would be included between two bulk electrodes, the resistance of the system would be proportional to the length of the wire. On the contrary, in the case of a one-dimensional (1D) *ballistic* quantum wire, the resistance equals $1/nG_0$, independently of the length of the wire. Here $n$ is the number of conduction channels ($n = 2$ for carbon nanotubes). With ballistic transport, conductance does not depend on the length of the 1D channel connecting bulk electrodes. Thus our observation of almost perfectly flat plateaus on the $G(x)$ curves provides evidence for quasi-ballistic transport in DWNTs at room temperature.

A $2G_0$ conductance is expected for an ideal metallic nanotube, which makes perfect contacts with both electrodes [16]. Although a series of steps was observed in most of the samples, the theoretically expected $2G_0$ conductance jumps were not found in our DWNT samples. The step size of $1G_0$ reported by Frank *et al*. [12] was not observed also. The height of the first conductance step was typically in the range $0.05G_0 - 0.15G_0$, which is much less than $2G_0$. To explain this deviation from the theoretically expected value, one has to take into account the contact effects. In our setup, a nanotube that touches the Hg electrode does not usually make direct contact to the metallic mobile probe. Instead, it connects to other tubes in the soot, which means the current flows from the probe to the measured DWNT through many tube-to-tube junctions. Thus the contact resistance between the probe and the measured tube is high and varies from sample to sample [17]. We believe that this high contact resistance reduces the height of the conductance jump below the theoretically expected value of $2G_0$ and causes it to vary from sample to sample.

In what follows we analyze the observed small deviations from the perfect horizontal orientation of the plateaus and use this information to calculate the electronic mean free path (EMFP) in DWNTs. Using the first plateau in the $G(x)$ curves, we estimated the resistance per unit length of our DWNTs. For this estimation, the conductance $G(x)$ was converted into resistance as $R(x) = 1/G(x)$ (Fig. 3B). Since Hg typically makes good electrical contact with MWNTs [12], we assume that the tube-Hg contact resistance is independent of the submerged length $x$. Thus total resistance $R$ can be expressed as $R(x) = R_c - \rho x$. Here $\rho$ is the resistance per unit length and $R_c$ is the contact resistance [15]. For an ideal tube with perfect contacts, one expects that $R(x) = (2G_0)^{-1} = const$, with $R(x)$ being independent of $x$. As shown in Fig. 3B, the best linear fit to the $R(x)$ curve gives us $R_c$ = 170 kΩ and $\rho$ = 4.5 kΩ/μm. The $R_c$ and $\rho$ parameters obtained from 47 traces measured on six different soot samples fall in the range 100 – 200 kΩ and 0.1 – 11 kΩ/μm respectively, with the mean values $R_c$ = 165 kΩ and $\rho$ = 3 kΩ/μm. The resistance per unit length ($\rho$) of the carbon nanotube is related to the EMFP ($l$) as $\rho = (h/4e^2)(1/l)$ [18]. This equation allows us to determine that about 80% of tubes have their EMFP in the range 0.6-10 μm. Figure 4 shows the distribution of the obtained EMFP, in which the maximum is found at ~1.5 μm.

To determine whether these results are consistent with the theoretically expected mean free path, we calculated the electron acoustic-phonon scattering rate [13] under the assumption that only the outer layer of the DWNTs contributes to the conduction process [12,19]. We also assume that the tube has a metallic armchair structure with a ($N_B$, $N_B$) chiral index. The electron–acoustic phonon scattering rate $1/\tau_{ac}$ can be expressed as $\frac{1}{\tau_{ac}} = \frac{8\pi^2}{h}\Xi^2(\frac{k_B T}{2\sigma v_s^2})\frac{1}{hv_F} = \frac{v_F}{l_{ac}}$, where $\Xi \approx 5 eV$ is the deformation potential (Ref. 13 and eq. (5) in Ref. 20), $k_B$ is Boltzmann's constant, $T$ is the temperature, σ is the nanotube mass per unit length, $v_s$ = 1.5 x 10$^4$ m/s is the acoustic phonon velocity, $v_F$ = 8 x 10$^5$ m/s is Fermi velocity, and $l_{ac}$ is the electron acoustic-phonon scattering mean free path [13]. To apply this formula, we use the diameter range 2-7 nm obtained from TEM imaging (this corresponds to chirality $N_B$ = 15-50, assuming the tubes are of armchair structure). We obtain $l_{ac}$ in the range 3 – 10 μm, which is in good agreement with the experimental values obtained from many of the tested nanotubes (Fig. 4). Yet, about half of the tested samples showed $l$ < 3 μm (Fig. 4). This deviation from the calculated $l_{ac}$ can be explained by the effect of disorder. In the disorder model, the mean free path ($l_d$) in an armchair tube can be expressed as $l_d = \dfrac{6V_0^2}{2\sigma_\varepsilon^2 + 9\sigma_V^2} N_B \times \dfrac{\sqrt{3}}{2} d_0$, where $V_0$ = -2.7eV is the nearest-neighbor C-C tight binding overlap energy and $d_0 = 0.142$ nm is the C-C bond distance [6]. In this model, the value of the diagonal and off-diagonal elements in the Hamiltonian matrix are not fixed as they are in the ideal tube but are independent random variables with variances $\sigma_\varepsilon^2$ and $\sigma_V^2$, respectively [6]. We consider a positionally disordered tube with a variable C-C bond length, which has $\sigma_\varepsilon$ = 0 eV and $\sigma_V \neq$ 0 eV. By choosing $\sigma_V$ = 0.12 eV, we obtain $l_d$ in the range 0.6 – 2.1 μm for armchair tubes with $N_B$ = 15 – 50, which agrees with the shortest mean free path values observed. Since the variation in the Hamiltonian elements can be described as $\delta V = \alpha \delta d$ with α = 47 eV/nm [6], $\sigma_V$ = 0.12 eV provides $\delta d$ = 2.6 x 10$^{-3}$ nm, suggesting that the C-C length varies by ~2%. In general, we conclude that both types of scattering, i.e. electron-phonon scattering and disorder induced scattering, are significant in our samples.

In summary, we measured the electrical resistance versus length dependence of double-wall carbon nanotubes by submerging them into liquid mercury. The experiment demonstrates quasi-ballistic transport in DWNTs, with a remarkably long electronic mean free path of a few microns at room temperature. The measured mean free path is in good agreement with the calculations based on the scattering by acoustic phonons and by disorder. The degree of positional disorder of the tube was found to be ~2%.

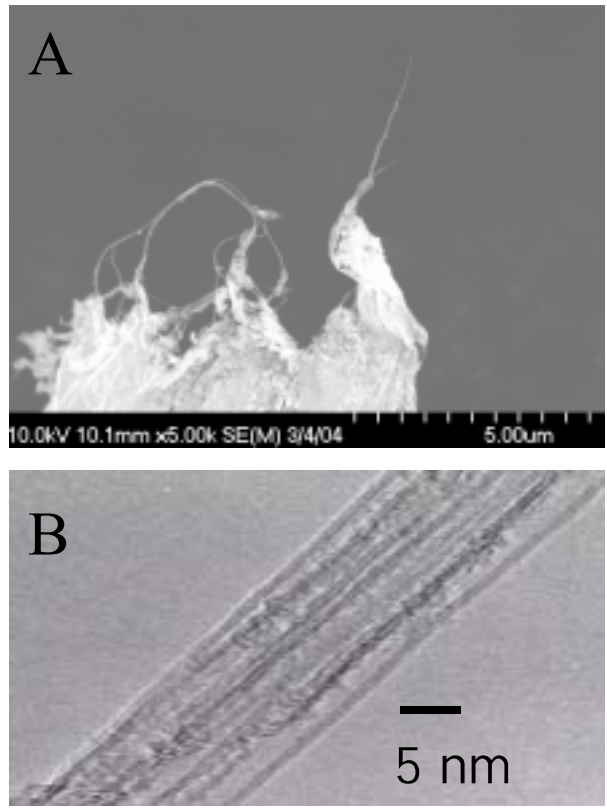

Fig. 1. (A) SEM image showing that DWNT bundles protruding from the edges of the threadlike soot sample. (B) TEM image showing the outer diameter of the DWNTs to range 2 – 7 nm.

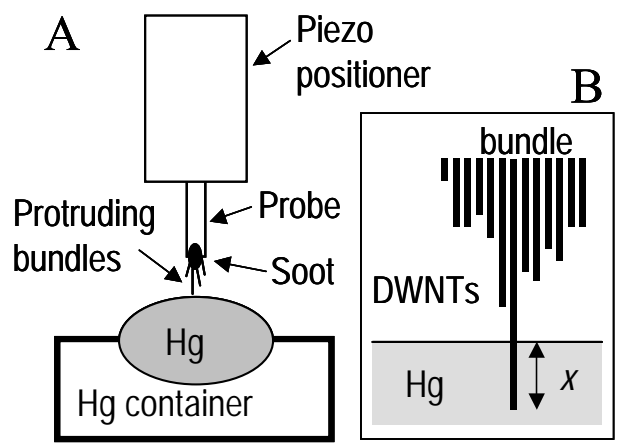

Fig. 2. Schematic diagrams of (A) experimental setup and (B) tube-Hg contact. Here $x$ is the length of the tube segment submerged into Hg.

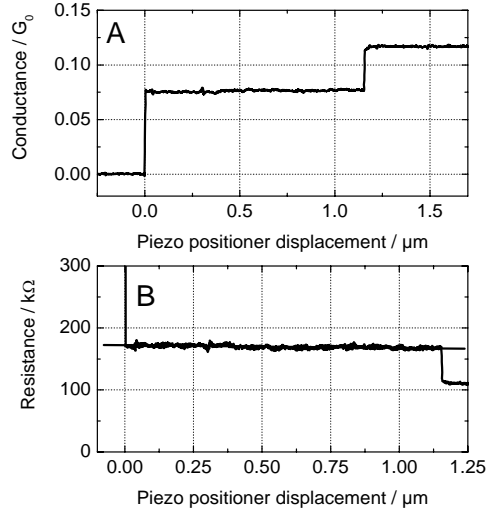

Fig. 3. (A) Conductance trace $G(x)$ normalized by $G_0 = 2e^2/h$ and (B) resistance trace $R(x) = 1/G(x)$ as a function of piezo positioner displacement ($x$) measured as the tube is pushed into Hg. The straight line in (B) is the linear fit given by $R(x) = C - \rho x$. Here C is one fitting parameter and $C \approx R_c$, and $\rho$ is the other fitting parameter representing the resistance per unit length of the nanotube. In this example $C = 170$ k$\Omega$ and $\rho = 4.5$ k$\Omega$/μm.

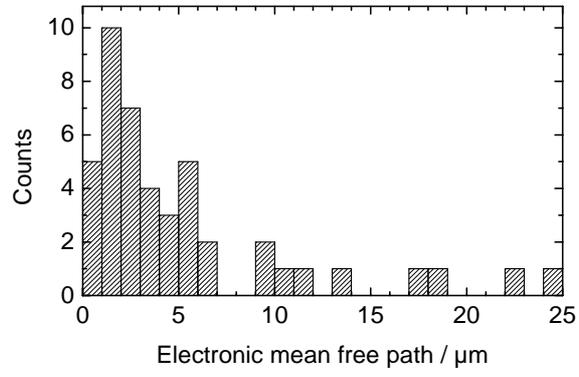

Fig. 4. Histogram of the electronic mean free path of the DWNTs.